%% file: main.tex
\pdfoutput=1

\documentclass[12pt,a4paper]{article}

\usepackage{ifthen} 
\newboolean{pdflatex}
\setboolean{pdflatex}{true} 

\newboolean{articletitles}
\setboolean{articletitles}{true} 

\newboolean{uprightparticles}
\setboolean{uprightparticles}{false} 

\input{preamble}

\begin{document}

\renewcommand{\thefootnote}{\fnsymbol{footnote}}
\setcounter{footnote}{1}

\input{title-LHCb-PAPER}

\renewcommand{\thefootnote}{\arabic{footnote}}
\setcounter{footnote}{0}

\pagestyle{plain} 
\setcounter{page}{1}
\pagenumbering{arabic}

\input{introduction}

\input{detector}
\input{dataset_and_selection}

\input{yields}

\input{systematics}

\input{results}

\input{acknowledgements}

\addcontentsline{toc}{section}{References}
\bibliographystyle{LHCb}
\bibliography{main,LHCb-PAPER,LHCb-CONF}

\end{document}

%% file: preamble.tex
\usepackage{microtype}
\usepackage{lineno}  
\usepackage{xspace} 

\usepackage{graphicx}  
\usepackage{color}
\usepackage{colortbl}
\graphicspath{{./figs/}} 

\usepackage{amsmath} 
\usepackage{amssymb}
\usepackage{amsfonts}
\usepackage{upgreek} 

\newcommand*\patchAmsMathEnvironmentForLineno[1]{%
\expandafter\let\csname old#1\expandafter\endcsname\csname #1\endcsname
\expandafter\let\csname oldend#1\expandafter\endcsname\csname
end#1\endcsname
 \renewenvironment{#1}%
   {\linenomath\csname old#1\endcsname}%
   {\csname oldend#1\endcsname\endlinenomath}%
}
\newcommand*\patchBothAmsMathEnvironmentsForLineno[1]{%
  \patchAmsMathEnvironmentForLineno{#1}%
  \patchAmsMathEnvironmentForLineno{#1*}%
}
\AtBeginDocument{%
\patchBothAmsMathEnvironmentsForLineno{equation}%
\patchBothAmsMathEnvironmentsForLineno{align}%
\patchBothAmsMathEnvironmentsForLineno{flalign}%
\patchBothAmsMathEnvironmentsForLineno{alignat}%
\patchBothAmsMathEnvironmentsForLineno{gather}%
\patchBothAmsMathEnvironmentsForLineno{multline}%
}

\usepackage{hyperref}    
\usepackage[all]{hypcap} 

\input{lhcb-symbols-def} 

\usepackage{cite} 
\usepackage{mciteplus}

%% file: lhcb-symbols-def.tex
\def\lhcb {\mbox{LHCb}\xspace}
\def\ux85 {\mbox{UX85}\xspace}

\def\babar  {\mbox{BaBar}\xspace}

\def\cleoc   {\mbox{CLEO-c}\xspace}

\ifthenelse{\boolean{uprightparticles}}%
{

 \def\Ppi         {\ensuremath{\uppi}\xspace}

 \def\PDelta      {\ensuremath{\Delta}\xspace}                 
 \def\PXi      {\ensuremath{\Xi}\xspace}                 
 \def\PLambda      {\ensuremath{\Lambda}\xspace}                 
 \def\PSigma      {\ensuremath{\Sigma}\xspace}                 
 \def\POmega      {\ensuremath{\Omega}\xspace}                 
 \def\PUpsilon      {\ensuremath{\Upsilon}\xspace}                 
 
 \def\PB      {\ensuremath{\mathrm{B}}\xspace}                 
                  
 \def\PD      {\ensuremath{\mathrm{D}}\xspace}

 \def\PK      {\ensuremath{\mathrm{K}}\xspace}

 \def\Pb      {\ensuremath{\mathrm{b}}\xspace}                 
 \def\Pc      {\ensuremath{\mathrm{c}}\xspace}

 \def\Pi      {\ensuremath{\mathrm{i}}\xspace}

 \def\Ps      {\ensuremath{\mathrm{s}}\xspace}

}
{

 \def\Ppi         {\ensuremath{\pi}\xspace}

 \mathchardef\PDelta="7101
 \mathchardef\PXi="7104
 \mathchardef\PLambda="7103
 \mathchardef\PSigma="7106
 \mathchardef\POmega="710A
 \mathchardef\PUpsilon="7107
                  
 \def\PB      {\ensuremath{B}\xspace}                 
                  
 \def\PD      {\ensuremath{D}\xspace}

 \def\PK      {\ensuremath{K}\xspace}

 \def\Pb      {\ensuremath{b}\xspace}                 
 \def\Pc      {\ensuremath{c}\xspace}

 \def\Pi      {\ensuremath{i}\xspace}

 \def\Ps      {\ensuremath{s}\xspace}

}

\def\squark    {\ensuremath{\Ps}\xspace}

\def\cquark    {\ensuremath{\Pc}\xspace}

\def\bquark    {\ensuremath{\Pb}\xspace}

\def\pion  {\ensuremath{\Ppi}\xspace}
\def\piz   {\ensuremath{\pion^0}\xspace}

\def\pip   {\ensuremath{\pion^+}\xspace}
\def\pim   {\ensuremath{\pion^-}\xspace}

\def\pipm  {\ensuremath{\pion^\pm}\xspace}
\def\pimp  {\ensuremath{\pion^\mp}\xspace}

\def\kaon  {\ensuremath{\PK}\xspace}
  \def\Kbar  {\kern 0.2em\overline{\kern -0.2em \PK}{}\xspace}

\def\Kz    {\ensuremath{\kaon^0}\xspace}
\def\Kzb   {\ensuremath{\Kbar^0}\xspace}
\def\KzKzb {\ensuremath{\Kz \kern -0.16em \Kzb}\xspace}
\def\Kp    {\ensuremath{\kaon^+}\xspace}
\def\Km    {\ensuremath{\kaon^-}\xspace}

\def\KpKm  {\ensuremath{\Kp \kern -0.16em \Km}\xspace}
\def\KS    {\ensuremath{\kaon^0_{\rm\scriptscriptstyle S}}\xspace}

\def\Kstar   {\ensuremath{\kaon^*}\xspace}

  \def\Dbar    {\kern 0.2em\overline{\kern -0.2em \PD}{}\xspace}
\def\D       {\ensuremath{\PD}\xspace}

\def\Dz      {\ensuremath{\D^0}\xspace}
\def\Dzb     {\ensuremath{\Dbar^0}\xspace}
\def\DzDzb   {\ensuremath{\Dz {\kern -0.16em \Dzb}}\xspace}
\def\Dp      {\ensuremath{\D^+}\xspace}
\def\Dm      {\ensuremath{\D^-}\xspace}
\def\Dpm     {\ensuremath{\D^\pm}\xspace}

\def\DpDm    {\ensuremath{\Dp {\kern -0.16em \Dm}}\xspace}

\def\Dsp     {\ensuremath{\D^+_\squark}\xspace}
\def\DsDp     {\ensuremath{\D^+_{(\squark)}}\xspace}
\def\DsDpm     {\ensuremath{\D^{\pm}_{(\squark)}}\xspace}
\def\DsDm     {\ensuremath{\D^-_{(\squark)}}\xspace}

\def\Bbar    {\ensuremath{\kern 0.18em\overline{\kern -0.18em \PB}{}}\xspace}

  \def\Y#1S{\ensuremath{\PUpsilon{(#1S)}}\xspace}

\def\Lbar {\ensuremath{\kern 0.1em\overline{\kern -0.1em\PLambda}}\xspace}

\def\to                 {\ensuremath{\rightarrow}\xspace}

\def\CP                {\ensuremath{C\!P}\xspace}

\def\AT#1     {\ensuremath{A_{\mathrm{T}}^{#1}}\xspace}           

\def\C#1      {\ensuremath{\mathcal{C}_{#1}}\xspace}                       
\def\Cp#1     {\ensuremath{\mathcal{C}_{#1}^{'}}\xspace}                    
\def\Ceff#1   {\ensuremath{\mathcal{C}_{#1}^{\mathrm{(eff)}}}\xspace}        
\def\Cpeff#1  {\ensuremath{\mathcal{C}_{#1}^{'\mathrm{(eff)}}}\xspace}       
\def\Ope#1    {\ensuremath{\mathcal{O}_{#1}}\xspace}                       
\def\Opep#1   {\ensuremath{\mathcal{O}_{#1}^{'}}\xspace}                    

\newcommand{\tev}{\ensuremath{\mathrm{\,Te\kern -0.1em V}}\xspace}
\newcommand{\gev}{\ensuremath{\mathrm{\,Ge\kern -0.1em V}}\xspace}
\newcommand{\mev}{\ensuremath{\mathrm{\,Me\kern -0.1em V}}\xspace}
\newcommand{\kev}{\ensuremath{\mathrm{\,ke\kern -0.1em V}}\xspace}
\newcommand{\ev}{\ensuremath{\mathrm{\,e\kern -0.1em V}}\xspace}
\newcommand{\gevc}{\ensuremath{{\mathrm{\,Ge\kern -0.1em V\!/}c}}\xspace}
\newcommand{\mevc}{\ensuremath{{\mathrm{\,Me\kern -0.1em V\!/}c}}\xspace}
\newcommand{\gevcc}{\ensuremath{{\mathrm{\,Ge\kern -0.1em V\!/}c^2}}\xspace}
\newcommand{\gevgevcccc}{\ensuremath{{\mathrm{\,Ge\kern -0.1em V^2\!/}c^4}}\xspace}
\newcommand{\mevcc}{\ensuremath{{\mathrm{\,Me\kern -0.1em V\!/}c^2}}\xspace}

\def\mm   {\ensuremath{\rm \,mm}\xspace}

\def\mum  {\ensuremath{\,\upmu\rm m}\xspace}

\def\invfb   {\ensuremath{\mbox{\,fb}^{-1}}\xspace}

\def\ps   {\ensuremath{{\rm \,ps}}\xspace}

\def\gsim{{~\raise.15em\hbox{$>$}\kern-.85em
          \lower.35em\hbox{$\sim$}~}\xspace}
\def\lsim{{~\raise.15em\hbox{$<$}\kern-.85em
          \lower.35em\hbox{$\sim$}~}\xspace}

\def\pt         {\mbox{$p_{\rm T}$}\xspace}

\def\evtgen     {\mbox{\textsc{EvtGen}}\xspace}
\def\pythia     {\mbox{\textsc{Pythia}}\xspace}

\def\geant      {\mbox{\textsc{Geant4}}\xspace}

\def\tell1  {TELL1\xspace}
\def\ukl1   {UKL1\xspace}



%% file: title-LHCb-PAPER.tex
\begin{titlepage}
\pagenumbering{roman}

\vspace*{-1.5cm}
\centerline{\large EUROPEAN ORGANIZATION FOR NUCLEAR RESEARCH (CERN)}
\vspace*{1.5cm}
\hspace*{-0.5cm}
\begin{tabular*}{\linewidth}{lc@{\extracolsep{\fill}}r}
\ifthenelse{\boolean{pdflatex}}
{\vspace*{-2.7cm}\mbox{\!\!\!\includegraphics[width=.14\textwidth]{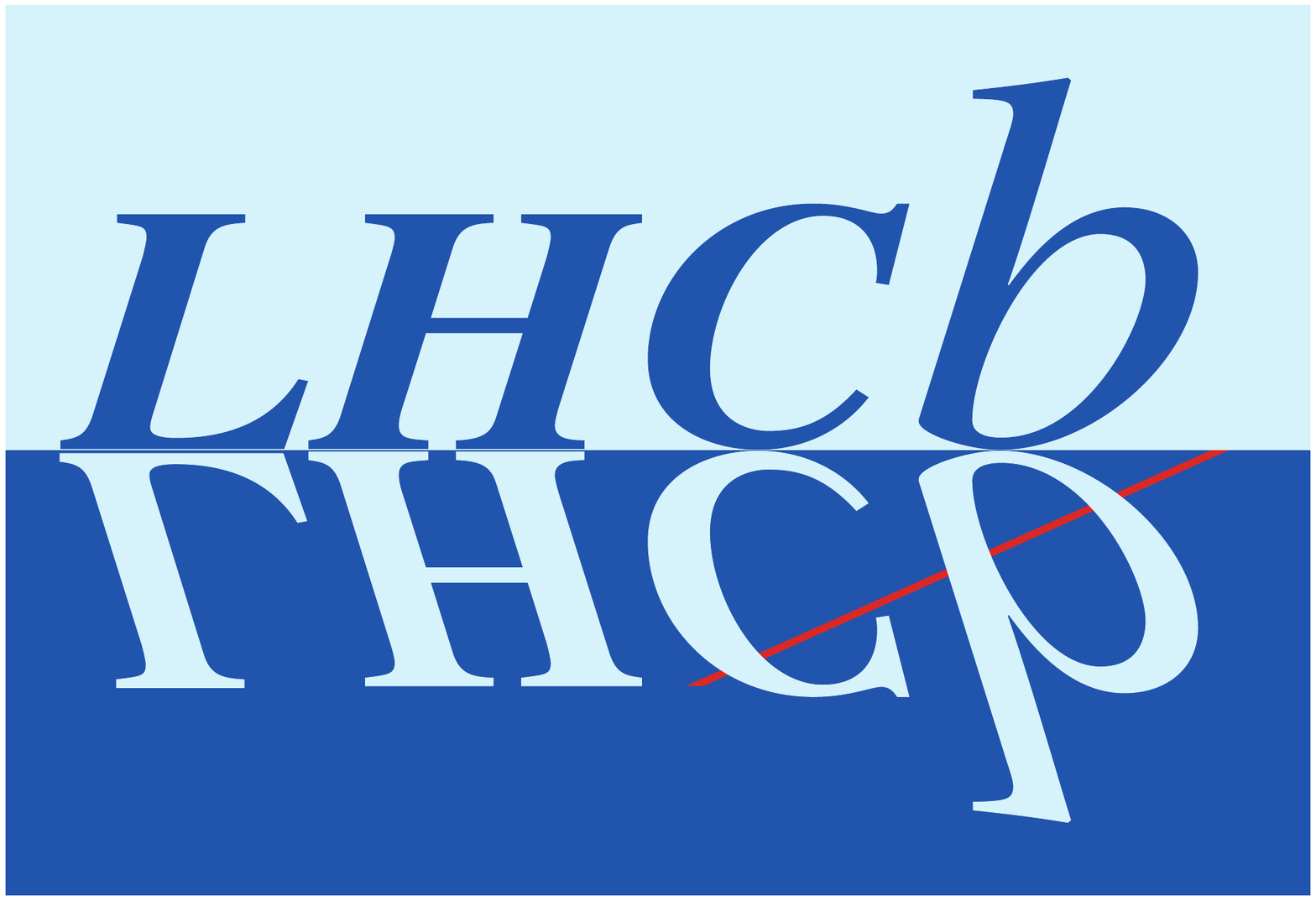}} & &}%
{\vspace*{-1.2cm}\mbox{\!\!\!\includegraphics[width=.12\textwidth]{lhcb-logo.eps}} & &}%
\\
 & & CERN-PH-EP-2013-021 \\  
 & & LHCb-PAPER-2012-052 \\  
 & & \today \\ 
 & & \\
\end{tabular*}

\vspace*{2.5cm}

{\bf\boldmath\huge
\begin{center}
  Search for \CP violation in $\Dp \to \phi\pip$ and $\Dsp \to
  \KS\pip$ decays
\end{center}
}

\vspace*{1.0cm}

\begin{center}
The LHCb collaboration\footnote{Authors are listed on the following pages.}
\end{center}
\vspace*{1.0cm}

\begin{abstract}
  \noindent
 A search for \CP violation in $\Dp \to \phi\pip$
 decays is performed using data collected in 2011 by the LHCb
 experiment corresponding to an integrated luminosity of 1.0\invfb at a
 centre of mass energy of 7\tev. The \CP-violating asymmetry is measured to be
 $(-0.04\pm0.14\pm0.14)\%$ for candidates with $\Km\Kp$
 mass within 20\mevcc of the $\phi$ meson mass. A search for a \CP-violating asymmetry
 that varies across the $\phi$ mass region of the $\Dp \to \Km\Kp\pip$
 Dalitz plot is also performed, and no evidence for \CP violation is
 found. In addition, the \CP asymmetry in the $\Dsp \to
 \KS\pip$ decay is measured to be
 $(0.61\pm0.83\pm0.14)\%$. 
\end{abstract}

\vspace*{2.0cm}

\begin{center}
  Published as JHEP06 (2013) 112
\end{center}

\vspace{\fill}

{\footnotesize 
\centerline{\copyright~CERN on behalf of the \lhcb collaboration, license \href{http://creativecommons.org/licenses/by/3.0/}{CC-BY-3.0}.}}
\vspace*{2mm}

\end{titlepage}

\newpage
\setcounter{page}{2}
\mbox{~}
\newpage

\input{LHCb_authorlist.tex}

\cleardoublepage

%% file: LHCb_authorlist.tex
\centerline{\large\bf LHCb collaboration}
\begin{flushleft}
\small
R.~Aaij$^{40}$, 
C.~Abellan~Beteta$^{35,n}$, 
B.~Adeva$^{36}$, 
M.~Adinolfi$^{45}$, 
C.~Adrover$^{6}$, 
A.~Affolder$^{51}$, 
Z.~Ajaltouni$^{5}$, 
J.~Albrecht$^{9}$, 
F.~Alessio$^{37}$, 
M.~Alexander$^{50}$
% ...

{\footnotesize \it
% ...
$ ^{5}$Clermont Universit\'{e}, Universit\'{e} Blaise Pascal, CNRS/IN2P3, LPC, Clermont-Ferrand, France\\
$ ^{6}$CPPM, Aix-Marseille Universit\'{e}, CNRS/IN2P3, Marseille, France\\
% ...
$ ^{9}$Fakult\"{a}t Physik, Technische Universit\"{a}t Dortmund, Dortmund, Germany\\
% ...
$ ^{35}$Universitat de Barcelona, Barcelona, Spain\\
$ ^{36}$Universidad de Santiago de Compostela, Santiago de Compostela, Spain\\
$ ^{37}$European Organization for Nuclear Research (CERN), Geneva, Switzerland\\
% ...
$ ^{40}$Nikhef National Institute for Subatomic Physics, Amsterdam, The Netherlands\\
% ...
$ ^{45}$H.H. Wills Physics Laboratory, University of Bristol, Bristol, United Kingdom\\
% ...
$ ^{50}$School of Physics and Astronomy, University of Glasgow, Glasgow, United Kingdom\\
$ ^{51}$Oliver Lodge Laboratory, University of Liverpool, Liverpool, United Kingdom\\
% ...
$ ^{n}$LIFAELS, La Salle, Universitat Ramon Llull, Barcelona, Spain\\
% ...
}
\end{flushleft}

%% file: introduction.tex
\section{Introduction}
\label{sec:intro}

Cabibbo-suppressed charm decays are the focus of searches for direct \CP violation (CPV) in the charm sector. In these decays, direct CPV will occur if tree and loop (penguin) processes interfere with different strong and weak phases. Furthermore, contributions from physics beyond the Standard Model may appear in the virtual loops~\cite{Grossman:2006jg}. Evidence for direct CPV in charm decays was reported by LHCb and subsequently by CDF using the $\Dz \to \Km\Kp$ and $\Dz \to \pim\pip$ channels~\cite{LHCb-PAPER-2011-023, Collaboration:2012qw}. Although the latest results do not confirm the evidence for CPV in the charm sector~\cite{Aaij:2013bra, LHCb-CONF-2013-003}, further studies using different decay modes remain well motivated. The large branching ratios of $\Dz \to \Km\Kp$ compared to $\Dz \to \pim\pip$ decays, and of the $\Dp \to \Km\Kp\pip$ compared to the $\Dp \to \pim\pip\pip$ mode, suggest that the contribution of the penguin amplitude may be significant in both $\Dz \to \Km\Kp$ and $\Dp \to \Km\Kp\pip$ decays ~\cite{Brod:2012ud}. The inclusion of charge conjugate decays is implied where appropriate throughout this paper. In \Dp decays, a non-zero \CP asymmetry would indicate unambiguously the presence of direct CPV. The $\Dp \to \phi\pip$ decay is a particularly promising channel for CPV searches due to its large branching ratio of $(2.65\pm0.09)\times10^{-3}$~\cite{PDG2012}. A recent investigation of this decay at the Belle experiment yielded a \CP-violating charge asymmetry of $(+0.51\pm0.28\pm0.05)\%$~\cite{Staric:2011en}, while \babar measured $(-0.3\pm0.3\pm0.5)\%$~\cite{Lees:2012nn}. 

Searches for CPV in charm decays with the LHCb experiment rely on a good understanding of the charge asymmetries both in $D$ meson production in $pp$ collisions and in the detection of the final states. These effects are studied using control decay modes in which no CPV is expected, and cancelled by measuring the differences in asymmetries between different final states or by comparing measurements made in one area of the Dalitz plot relative to another. 

To investigate CPV in the $\Dp \to \phi\pip$ decay, the $\Dp \to \KS\pip$ decay with $\KS \to \pim\pip$ is used as a control channel. This decay is itself sensitive to CPV via the interference of Cabibbo-favoured and doubly Cabibbo-suppressed amplitudes. However, the \CP asymmetry in this channel is predicted to be at most 0.01\% in the Standard Model~\cite{Bigi:1994aw}, and there is less scope for contributions from non-Standard Model dynamics than in the $\Dp \to \phi\pip$ decay as no penguin amplitudes contribute~\cite{Grossman:2006jg}. Therefore CPV in the $\Dp \to \KS\pip$ decay is assumed to be negligible. The \CP asymmetry in the $\Dp \to \phi\pip$ region of the $\Dp \to \Km\Kp\pip$ Dalitz plot is given by, to first order,
\begin{equation}
A_{CP}(\Dp \to \phi\pip)  = A_{\rm raw}(\Dp \to \phi\pip) - A_{\rm raw}(\Dp \to \KS\pip) + A_{CP}(\Kz/\Kzb),
\label{eq:acp}
\end{equation}
where the raw charge asymmetry $A_{\rm raw}$ is defined as 
\begin{equation}
A_{\rm raw} = \frac{N_{\Dp}-N_{\Dm}}{N_{\Dp}+N_{\Dm}},
\end{equation}
for yields $N_{\Dpm}$ of positively- or negatively-charged signal or control-mode candidates. The kaon asymmetry $A_{CP}(\Kz/\Kzb)$ is the correction for CPV in the neutral kaon system and is $-0.028\%$ with a systematic uncertainty of 0.028\%~\cite{LHCb:2012fb}. To first order, the use of the control channel cancels the effects of the \Dpm production asymmetry of $(-0.96\pm0.26\pm0.18)\%$~\cite{LHCb:2012fb} and of any asymmetry associated with the detection of the pion~\cite{Aaij:2012cy}. In the proximity of the $\phi$ meson mass of $1019.46\pm0.02$\mevcc~\cite{PDG2012} in the $\Dp \to \Km\Kp\pip$ Dalitz plot, the kaons have almost identical momentum distributions. Therefore the kaon interaction asymmetry cancels between the \Kp and \Km meson daughters of the $\phi$ resonance. Hence the search is restricted to decays with $\Kp\Km$ invariant masses in the range $1.00 < m_{\Km\Kp} < 1.04$\gevcc. 

A concurrent measurement of the \CP asymmetry in the $\Dsp \to \KS\pip$ decay, approximated as
\begin{equation}
A_{CP}(\Dsp \to \KS\pip)  = A_{\rm raw}(\Dsp \to \KS\pip) - A_{\rm raw}(\Dsp \to \phi\pip) +
A_{CP}(\Kz/\Kzb),
\label{eq:acpds}
\end{equation}
is performed using the $\Dsp \to \phi\pip$ decay as a control channel. This decay is also Cabibbo-suppressed, with similar contributions from loop amplitudes as the $\Dp \to \phi\pip$ decay, but the number of signal candidates is substantially lower. This is partly due to the lower \Dsp production cross-section~\cite{Aaij:2013mga} and partly because only \KS mesons with decay times of less than 40\ps are used in this analysis. In Eq.~(\ref{eq:acpds}), the effect of the CPV in the neutral kaon system has a sign opposite to that in Eq.~(\ref{eq:acp}) relative to the raw asymmetry in the $\DsDp \to \KS\pip$ decay because the \Dsp decays predominantly to a $\Kz$ meson while the \Dp decays to a $\Kzb$.

\begin{figure}
\begin{center}
\includegraphics[width=9cm]{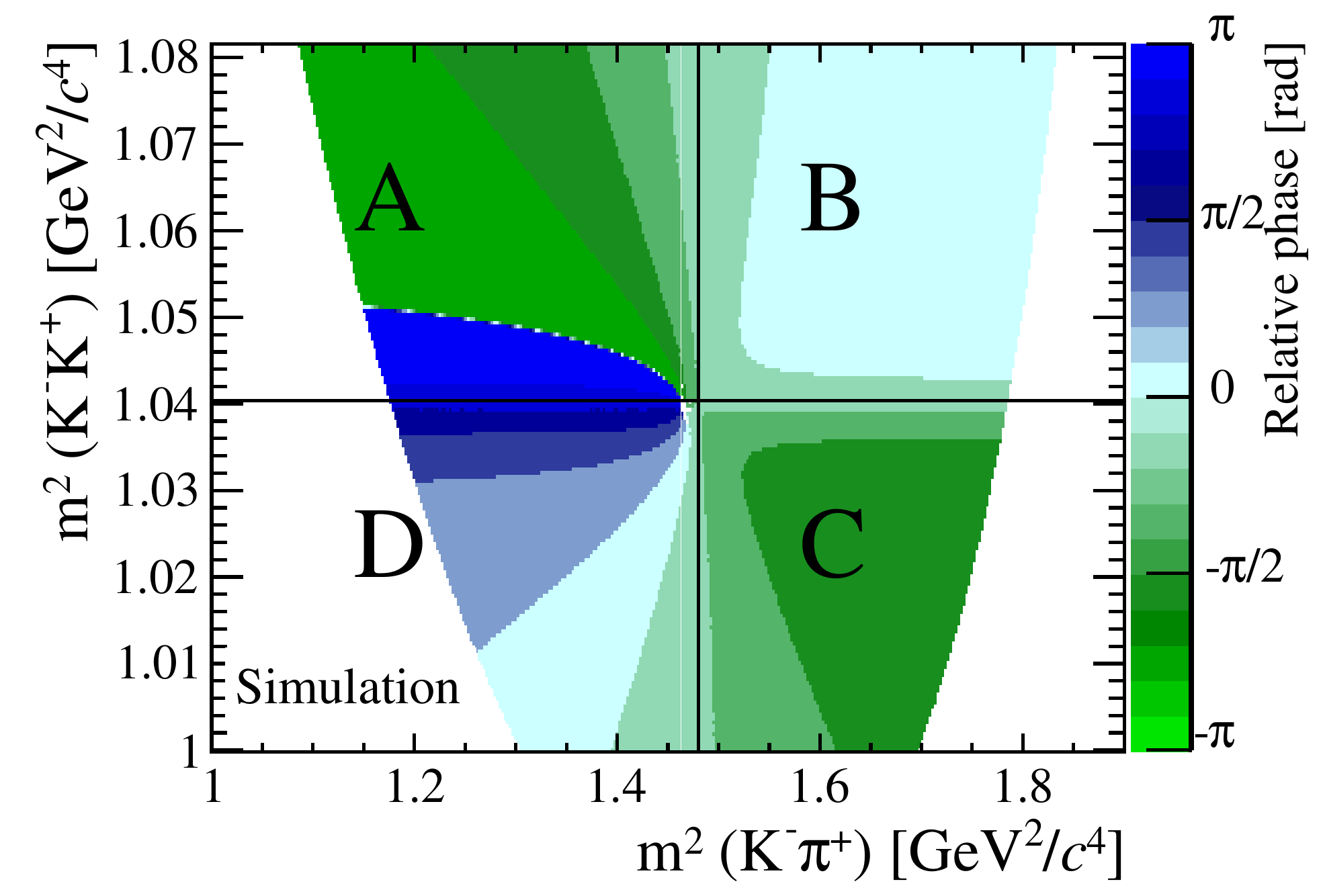}
\caption{Variation of the overall phase of the \Dp decay
  amplitude in the $\phi$ mass region of the Dalitz
  plot, from a simulation study based on the \cleoc amplitude model in which the phase is
  defined relative to that of the $\Kstar(892)^{0}$ resonance~\cite{:2008zi}. To calculate $A_{CP}|_{S}$, the region is divided into rectangular zones as shown, corresponding to $1.00 < m(\Km\Kp)< 1.02$\gevcc and $1.02 < m(\Km\Kp)< 1.04$\gevcc along the $y$-axis, and to $m^{2}(\Km\pip) < 1.48$\gevgevcccc and $m^{2}(\Km\pip) > 1.48$\gevgevcccc along the $x$-axis.\label{fig:strongphase}}
\end{center}
\end{figure}

Within the Standard Model, CPV in singly Cabibbo-suppressed charm decays with contributing tree and penguin amplitudes is expected to be~\cite{soniAtwoodCPV}
\begin{equation}
A_{CP} \approx \left|\mathit{Im}\left(\frac{V_{ub}V_{cb}^{*}}{V_{us}V_{cs}^{*}}\right)\right|R\sin\delta_{S},
\end{equation}
where $R$ is a number of order one that depends on hadronic matrix elements,
$\delta_{S}$ is the strong phase difference between tree and penguin amplitudes,
and $V_{ij}$ are elements of the CKM matrix. In the region of the $\phi$ resonance in the $\Dp \to \Km\Kp\pip$ Dalitz plot, several other amplitudes contribute to the overall matrix element and interfere with the $\phi$ meson~\cite{Lees:2012nn, :2008zi}. A recent amplitude analysis of this decay mode from the \cleoc collaboration~\cite{:2008zi} yields a matrix element with a relative strong phase that varies rapidly across the $\phi$ region, as shown in Fig.~\ref{fig:strongphase}. The isobar amplitude model favoured by \cleoc (fit `B' in Ref.~\cite{:2008zi}) contains major contributions from the $\phi$, $\Kstar(892)^{0}$, $\Kstar_{0}(1430)^{0}$ and $K^{*}_{0}(800)$ resonances. The phase is measured relative to that of the $\Kstar(892)^{0}$ meson.  The variation in phase means that it is possible that a constant \CP-violating asymmetry could be cancelled out when the different regions of the $\phi$ resonance are combined to calculate $A_{CP}$. Hence we define a complementary observable called $A_{CP}|_{S}$. The area around the $\phi$ resonance in the Dalitz plot is split into four rectangular regions $A-D$ defined clockwise from the top-left as shown in Fig.~\ref{fig:strongphase}. The division is chosen to minimise the change in phase within each region. A difference between the two diagonals, each made of two regions with similar phases, is calculated as
\begin{equation}
A_{CP}|_{S} =\frac{1}{2}\left(
 A_{\rm raw}^{A}+A_{\rm raw}^{C}-
 A_{\rm raw}^{B}-A_{\rm raw}^{D}\right).
\label{eq:acps}
\end{equation} 
This observable is not affected by the \Dpm production asymmetry and is robust against systematic biases from the detector.

\begin{table}
\caption{Expected mean values of $A_{CP}$ and $A_{CP}|_{S}$ for different types of \CP violation
  introduced into the simulated Dalitz plots, together with the significance with which a signal could be observed given estimated overall uncertainties in $A_{CP}$ and $A_{CP}|_{S}$ of 0.2\%.
\label{tab:sens}}
\begin{center}
\begin{tabular}{lcc}
Type of CPV & Mean $A_{CP}$ (\%) & Mean $A_{CP}|_{S}$ (\%) \\
\hline
$3^{\circ}$ in $\phi$ phase & $-0.01 \; (0.1\sigma)$ & $-1.02 \; (5.1\sigma)$\\
$0.8\%$ in $\phi$ amplitude & $-0.50 \; (2.5\sigma)$ & $-0.02 \; (0.1\sigma)$ \\
$4^{\circ}$ in $K^{*}_{0}(1430)^{0}$ phase & $\phantom{-}0.52 \; (2.6\sigma)$ & $-0.89 \; (4.5\sigma)$ \\
$4^{\circ}$ in $K^{*}_{0}(800)$ phase & $\phantom{-}0.70\; (3.5\sigma)$ & $\phantom{-}0.10 \; (0.5\sigma)$ \\
\end{tabular}
\end{center}

\end{table}

To test the hypothesis that $A_{CP}|_{S}$ can sometimes be more sensitive to \CP
violation than $A_{CP}$, a study is performed using simulated pseudo-experiments in which plausible types of CPV are introduced into the \cleoc
amplitude model~\cite{:2008zi}. The matrix elements for \Dp and \Dm
decays are separately modified in a number of ways, as specified in
Table~\ref{tab:sens}, and events are generated from the resulting probability density
functions. In each simulated sample, approximately the same number of events as in the
dataset are produced, and the values of $A_{CP}$ and
$A_{CP}|_{S}$ are compared. The effects of background and of the reconstruction and signal selection
efficiency variation across the $\phi$ region are negligible.

The level of CPV in the
 pseudo-experiments is chosen to give an expected result with significance
of around three Gaussian standard deviations in at least one observable. For each type of CPV, twenty Dalitz plots are simulated. The
mean values of $A_{CP}$ and $A_{CP}|_{S}$ in these pseudo-experiments are given
in Table~\ref{tab:sens}, together with the significance with which
these signals could be observed in the dataset under study. The table indicates that some types of CPV
can be observed more effectively with $A_{CP}$ and others with
$A_{CP}|_{S}$.

It was found in Ref.~\cite{Aaij:2011cw} that the sensitivity to CPV can vary substantially with the
details of the amplitude model. Therefore these simple simulations
should not be treated as accurate predictions, but instead as a
guide to the relative sensitivity of the two observables.

%% file: detector.tex
\section{Detector}
\label{sec:Detector}

The \lhcb detector~\cite{Alves:2008zz} is a single-arm forward
spectrometer covering the \mbox{pseudorapidity} range $2<\eta <5$,
designed for the study of particles containing \bquark or \cquark
quarks. The detector includes a high precision tracking system
consisting of a silicon-strip vertex detector (VELO) surrounding the $pp$
interaction region, a large-area silicon-strip detector located
upstream of a dipole magnet with a bending power of about
$4{\rm\,Tm}$, and three stations of silicon-strip detectors and straw
drift tubes placed downstream. The combined tracking system has
momentum resolution $\Delta p/p$ that varies from 0.4\% at 5\gevc to
0.6\% at 100\gevc, and impact parameter resolution of 20\mum for
tracks with high transverse momentum \pt. Charged hadrons are identified
using two ring-imaging Cherenkov detectors. Photon, electron and
hadron candidates are identified by a calorimeter system consisting of
scintillating-pad and preshower detectors, an electromagnetic
calorimeter and a hadronic calorimeter. Muons are identified by a
system composed of alternating layers of iron and multiwire
proportional chambers. The trigger~\cite{Aaij:2012me} consists of a hardware stage, based
on information from the calorimeter and muon systems, an
inclusive software stage, which uses the tracking
system, and a second software stage that exploits the full event
information. 

%% file: dataset_and_selection.tex
\section{Dataset and selection}
\label{sec:data}

The data sample used in this analysis corresponds to an integrated
luminosity of 1.0\invfb of $pp$
collisions at a centre of mass energy of $7$\tev, and was collected by
the LHCb experiment in 2011. The polarity of the LHCb
magnet was changed several times during the run, and
approximately half of the data were taken with each polarity, referred
to as `magnet-up' and `magnet-down' data hereafter.  To optimise the event selection and estimate
background contributions, 12.5 million $pp$ collision events
containing $\Dp \to \KS\pip$, $\KS \to \pim\pip$ decays and 5 million events containing $\Dp \to \Km\Kp\pip$ decays are simulated
with \pythia~6.4~\cite{Sjostrand:2006za} with a specific \lhcb
configuration~\cite{LHCb-PROC-2010-056}. Hadron decays are described by \evtgen~\cite{Lange:2001uf}. The interaction of the generated particles with the detector and its response are implemented using the \geant toolkit~\cite{Allison:2006ve, *Agostinelli:2002hh} as described in Ref.~\cite{LHCb-PROC-2011-006}.

To ensure the dataset is unbiased, the trigger must accept candidates in well-defined
ways that can be shown to be charge-symmetric. A trigger decision may be based on part or all of
the \DsDp signal candidate, on other particles in the event, or both. For example, signal decays triggered at the
hardware level exclusively by the pion from the \DsDp decay are not used, as they are shown in Sect.~\ref{sec:sys}
to have large detector-dependent charge asymmetries. 
For an event to be accepted by the hardware trigger, two criteria, not mutually exclusive, are
used: the decision must be based on one of the daughter tracks of
the \KS or $\phi$ meson, or on a particle other than the decay products of the
\DsDp candidate. In the first case the same track must also activate the
inclusive software trigger. This software trigger requires that one of the tracks from
the signal \DsDp candidate has \pt $> 1.7$\gevc
and distance of closest approach to the primary vertex (PV) of at least 0.1\mm.  The
second stage of the software trigger is required to 
find combinations of three tracks that meet the criteria to be signal
decays. 

Candidate $\DsDp \to \phi\pip$ decays are reconstructed by combining the tracks from two
oppositely charged particles that are identified by the RICH detectors
as kaons with one track identified as a pion. The combined invariant mass of the two
kaons is required to lie in the range $1.00 < m_{\Km\Kp} < 1.04$\gevcc. The
scalar sum of the \pt of the daughter particles must exceed 2.8\gevc. 

To reconstruct $\DsDp \to \KS\pip$ candidates, pairs of oppositely
charged particles with a pion mass hypothesis are combined to form \KS
candidates. Only those with $\pt > 700$\mevc and invariant
mass within 35\mevcc of the world average \KS mass~\cite{PDG2012} are retained. Accepted
candidates are then combined with a third charged particle, the bachelor
pion, to form a \DsDp
candidate. The mass of the \KS meson is constrained to its known value
in the kinematic fit. All three pion tracks must
be detected in the VELO, so only \KS mesons with short decay times are used. 

Further requirements are applied in order to reduce
background from random track combinations and partially reconstructed
charm and $B$ decays. Both \KS and \DsDp candidates are required
to have a vertex with acceptable fit quality.
Daughters of the $\phi$ and $\KS$ mesons must have momentum $p > 2$\gevc
and $\pt > 250$\mevc. Impact parameter requirements are used to
ensure that all the daughters of the \DsDp candidate do not originate at
any PV in the event. To remove non-resonant $\Dp \to \pim\pip\pip$ candidates,
the \KS meson decay vertex must be displaced by at least 10\mm in the
forward direction from the
decay vertex of its parent \Dp meson.
The bachelor pion in both final states must
have $p > 5$\gevc and $\pt > 500$\mevc, must not come from
any PV, and must be positively
identified as a pion rather than as a kaon, electron or
muon. 
In addition, fiducial requirements are 
applied~\cite{LHCb-PAPER-2011-023} to exclude regions with a large 
tracking efficiency asymmetry. 
The \DsDp candidate is required to have $1.5 < \pt < 20.0$\gevc and pseudorapidity $\eta$ in the range $2.2 < \eta < 4.4$, 
to point to a PV
(to suppress $D$ from $B$ decays), and to have a decay time
significantly greater than zero. The proportion of events with more
than one \DsDp candidate is negligible.

\begin{figure}
\begin{center}
\includegraphics*[width=0.45\columnwidth]{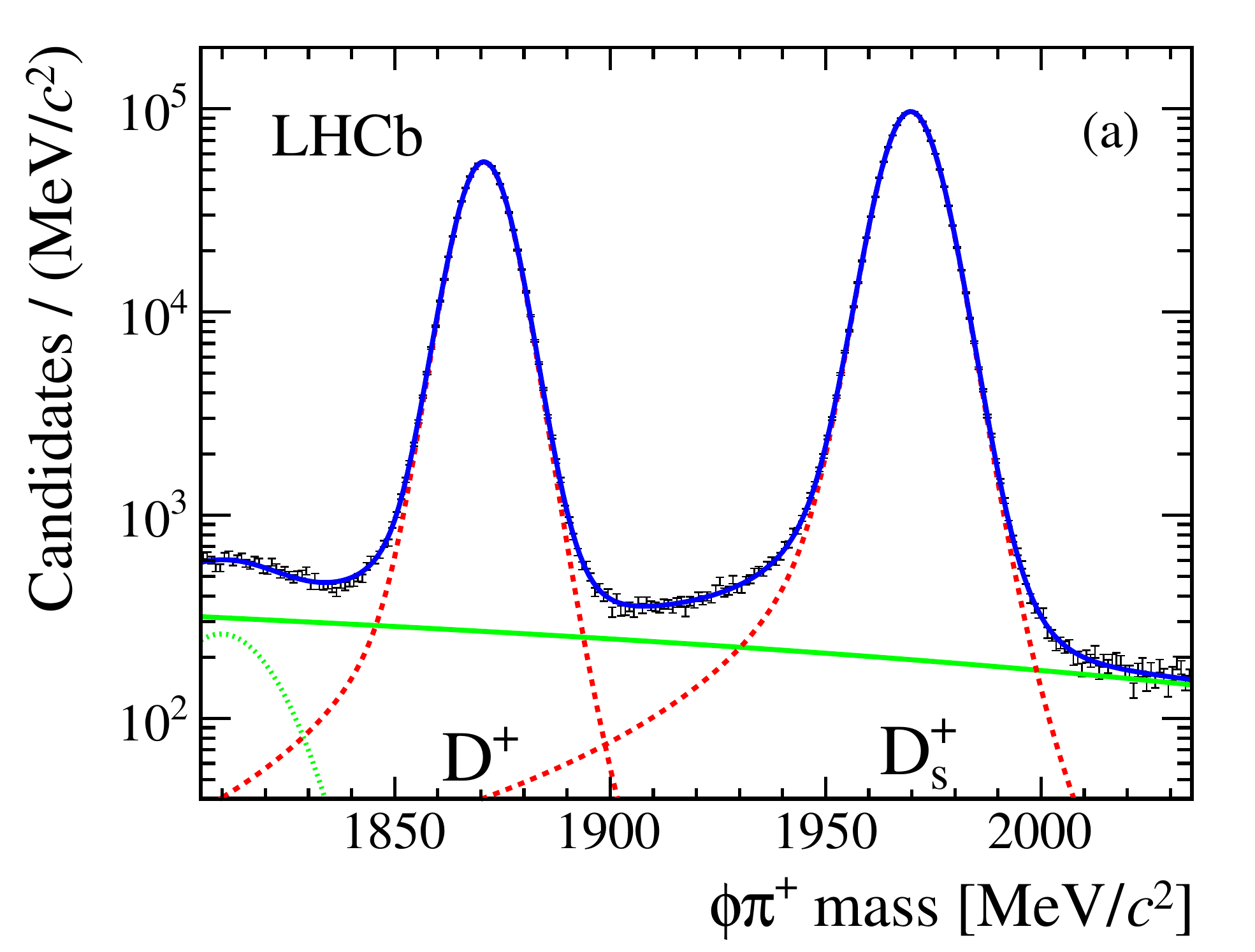}
\includegraphics*[width=0.45\columnwidth]{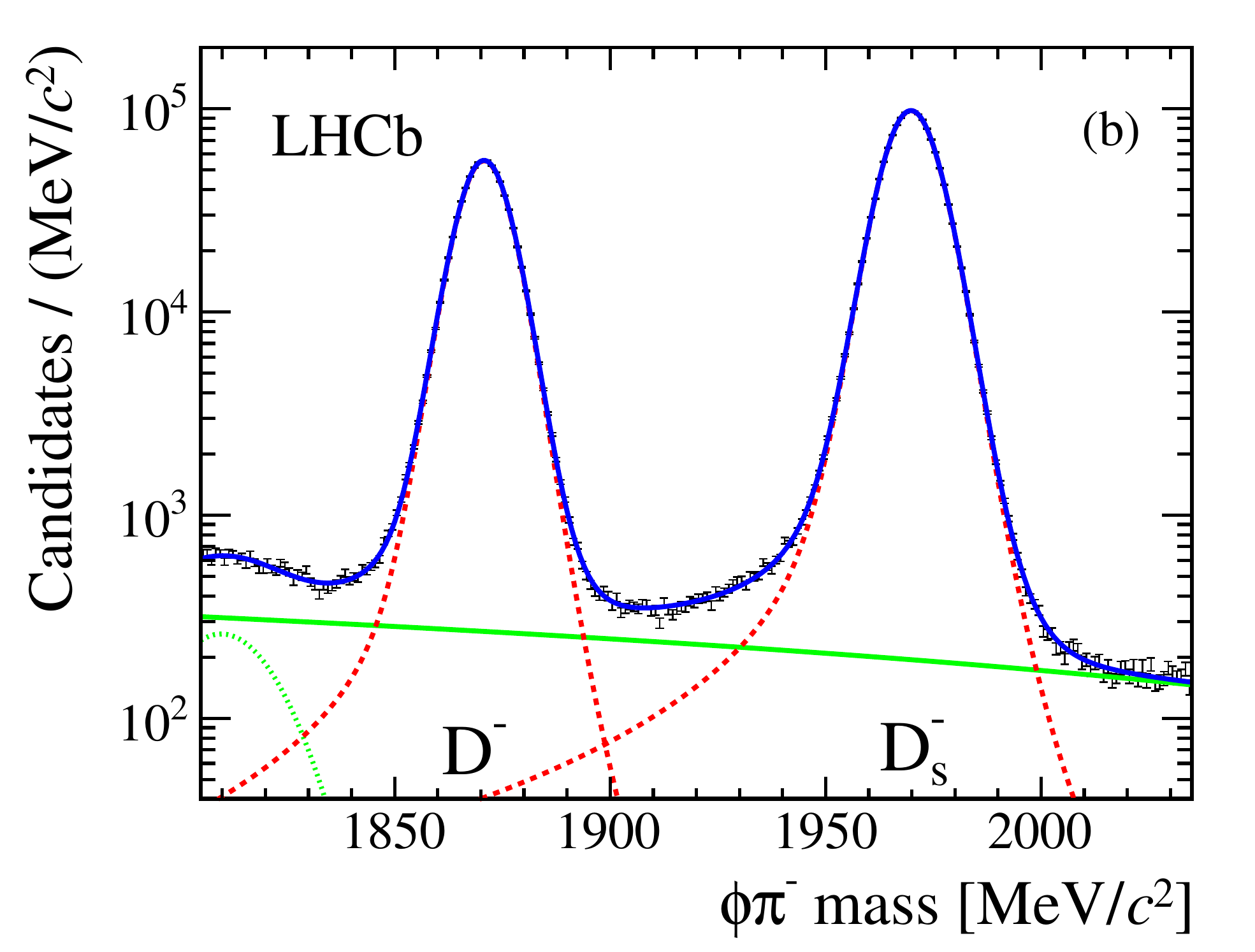}

\includegraphics*[width=0.45\columnwidth]{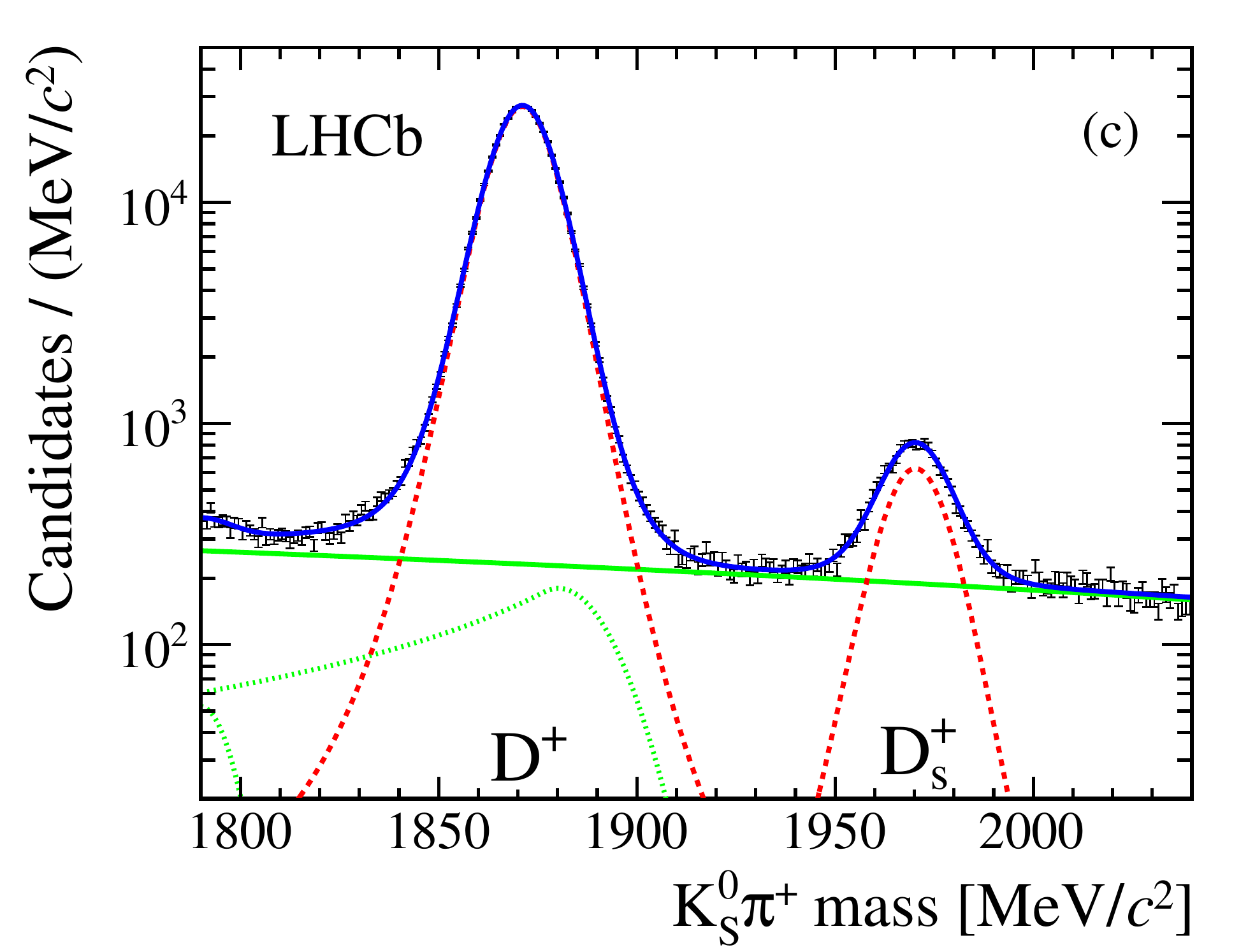}
\includegraphics*[width=0.45\columnwidth]{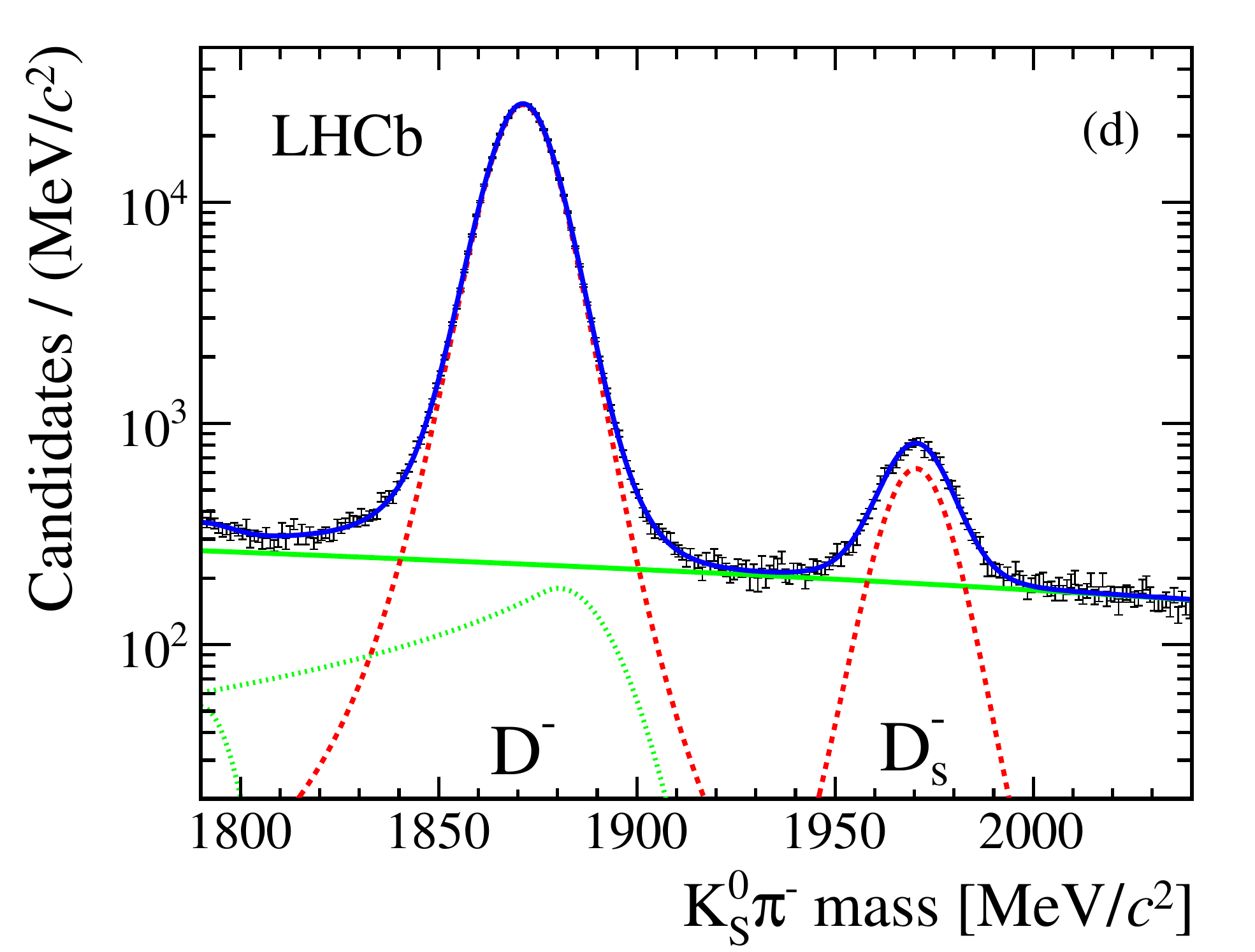}
\end{center}
\caption{
  Invariant mass distribution of selected (a) $\Dp \to \phi\pip$, 
  (b) $\Dm \to \phi\pim$, (c) $\Dp \to \KS\pip$ and (d) $\Dm \to
  \KS\pim$ candidates. The data are
  represented by symbols with error bars. The red dashed curves
  indicate the signal lineshapes,
  the green solid lines represent the combinatorial background shape, and
  the green dotted lines represent background from mis-reconstructed
 $\Dsp \to \phi\pip\piz$ decays in (a) and (b), and $\Dsp \to \KS\pip\piz$
  or $\Dsp \to \KS\Kp$ decays in (c) and (d). The blue solid lines show the sum of all fit components.
}
\label{fig:yields}
\end{figure}

\begin{figure}
\begin{center}
\includegraphics[width=9cm]{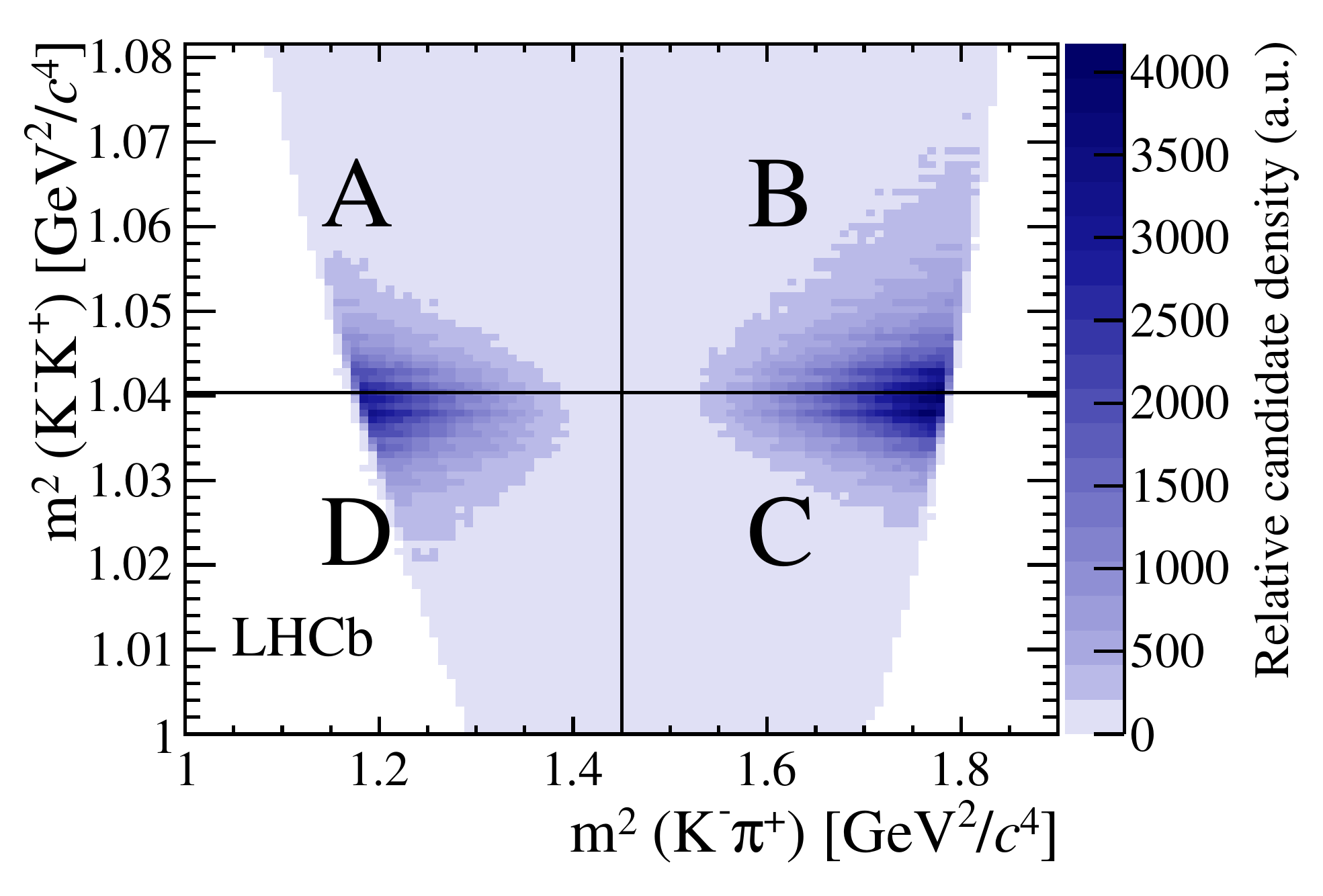}
\caption{Observed density of decays
  in the $\Dp \to \Km\Kp\pip$ Dalitz plot, with the regions A-D
  labelled as described in the text. \label{fig:phiregion}}
\end{center}
\end{figure}

The invariant mass distributions of
selected candidates in the two final states are presented in Fig.~\ref{fig:yields}. After applying
the selection and trigger requirements, 1,203,930 $\DsDpm \to \KS\pipm$
and 4,704,810 $\DsDpm \to \phi\pipm$ candidates remain in the mass ranges shown in
the figure. The distribution of
decays in the $\phi$ region of the $\Dp \to \Km\Kp\pip$
Dalitz plot is shown in Fig.~\ref{fig:phiregion}.

%% file: yields.tex
\section{Determination of the yields and asymmetries}
\label{sec:yields}

\begin{figure}
\begin{center}
\includegraphics[width=7cm]{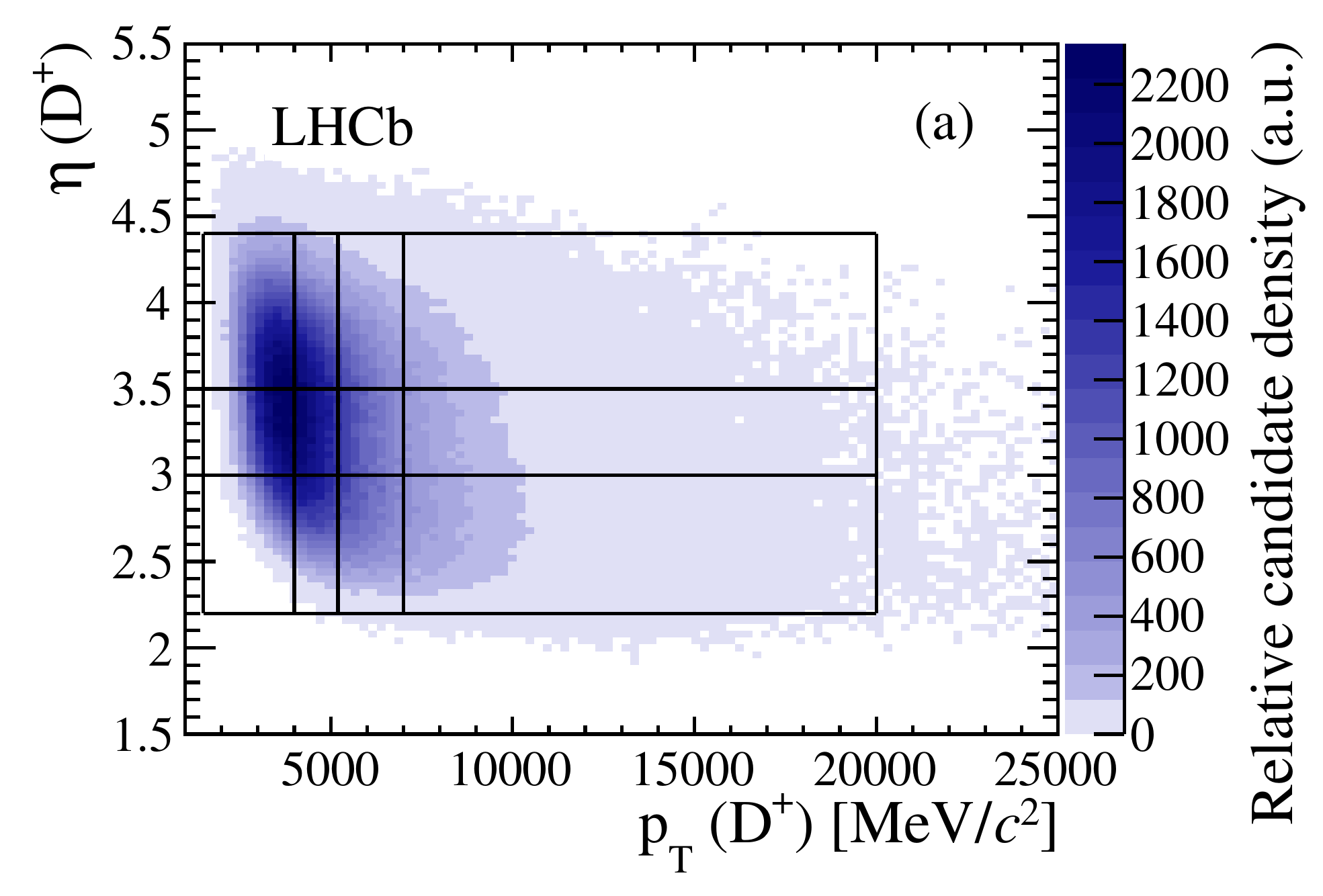}
\includegraphics[width=7cm]{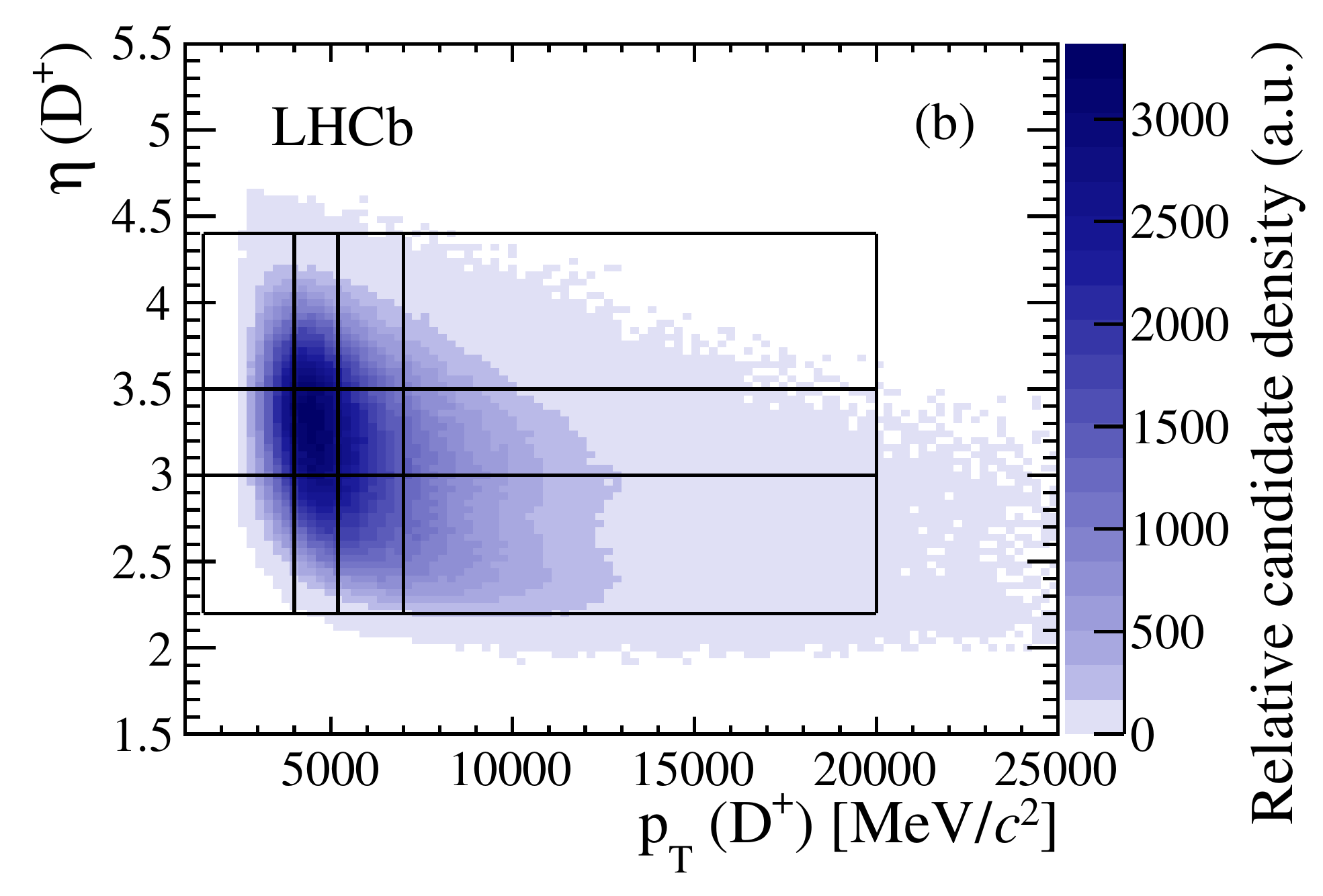}

\caption{Distributions of transverse momentum \pt and pseudorapidity
  $\eta$ for (a) $\Dp \to \KS\pip$ and (b) $\Dp \to \phi\pip$
 candidates with invariant masses $m$ in the range $1845 < m < 1895$\mevcc. Candidates
  that do not fall into the 12 rectangular bins are not used in the
  analysis.\label{fig:pteta} }
\end{center}
\end{figure}

For the measurement of $A_{CP}$, the signal yields are measured in 12
bins of transverse momentum \pt and pseudorapidity $\eta$, using binned
likelihood fits to the distributions of the invariant masses $m$, where
$m$ is either $m_{\phi\pip}$ or $m_{\KS\pip}$. The
values of $A_{CP}$ in each bin are calculated and a weighted average
over the bins is performed to obtain the final result. This procedure
is adopted because the distributions of the two decays in $\pt$ and
$\eta$ differ slightly, as shown in Fig.~\ref{fig:pteta}, and the \Dpm production asymmetry may also vary
over this range~\cite{LHCb:2012fb}. The $\pt-\eta$ binning therefore
reduces a potential source of systematic bias. The shapes of the $D^{+}_{(s)} \to
\KS\pip$ mass peaks are described by single Cruijff functions
\cite{delAmoSanchez:2010ae},
\begin{equation}
f(m) \propto \exp\left[\frac{-(m-\mu)^2}{2\sigma^{2}+(m-\mu)^2\alpha_{L,R}}\right]
\label{eq:cruijff}
\end{equation}
with the peak position defined by the free parameter $\mu$, the width by $\sigma$, and the tails by $\alpha_{L}$ and $\alpha_{R}$. The parameter
$\alpha_{L}$ is used for $m < \mu$ and
$\alpha_{R}$ for $m > \mu$. In the $\phi\pip$ final state,
Crystal Ball functions~\cite{Skwarnicki:1986xj} are added to the Cruijff functions to account
for the tails of the mass peaks. The signal lineshapes are tested on
simulated data and found to describe the data well. 
The background is fitted with a straight line and an additional
Gaussian component centred at low mass to account
for partially reconstructed $\Dsp \to \KS(\phi)\pip\piz$ decays. This
background mostly lies outside the interval in invariant mass that is fitted.
In the $\KS\pip$ case there is also a cross-feed component from
the $\Dsp \to \KS\Kp$ decay mode, where the $\Kp$ meson is misidentified
as a pion. In the
fit to data, the cross-feed yield and charge asymmetry are allowed to
vary but the shape is fixed from the simulation. It is modelled by a Crystal Ball function. The yield
of cross-feed is found to be small, at $6014\pm817$ decays, or
0.57\% of the \Dp yield. 

The fits are performed
simultaneously over four subsamples (\DsDp magnet-up, \DsDp magnet-down,
\DsDm magnet-up, and \DsDm magnet-down data) with the peak positions, widths and yields of
the \DsDp and background allowed to vary independently
in the four subsamples. All other parameters are shared. The
peak positions are found to differ between
charges and magnet polarities by around $0.2$\mevcc. The raw
asymmetries are then determined from the yields. The fitted yields are
given in in Table~\ref{tab:yields:yields}.

\begin{table}
\caption{Numbers of signal candidates in the four decay modes from
  the mass fits, with statistical uncertainties only. \label{tab:yields:yields}}
\begin{center}
\begin{tabular}{lcc}
Decay mode & Signal yield $(\times 10^{3})$\\
\hline
$\Dp \to \phi\pip$ &$1576.9\pm1.5$\\
$\Dsp \to \phi\pip$ & $3010.2\pm2.2$\\
$\Dp \to \KS\pip$ & $1057.8\pm1.2$\\
$\Dsp \to \KS\pip$ & $\phantom{00}25.6\pm0.2$\\
\end{tabular}
\end{center}

\end{table}

 The results are
cross-checked with a sideband subtraction procedure under the
assumption of a linear background. The background is sufficiently small relative
to the signal in the
$\Dp \to \phi\pip$ channel that the charge asymmetry can be calculated
by counting \Dp and \Dm candidates in a mass interval defined
around the \Dp mass of $1845< m < 1895$\mevcc. Therefore, the yields
for $A_{CP}|_{S}$ are evaluated using this simple
technique. The
resolution in the Dalitz plot is improved by
constraining the \Dp candidate mass to the world average
value~\cite{PDG2012}, instead of leaving it as a free parameter. This has a small effect which is assigned as a systematic uncertainty in
Sect.~\ref{sec:sys}. In the measurement of $A_{CP}$, the background
in the $\Dp \to \KS\pip$ channel is larger and therefore the results
are taken from fits.

%% file: systematics.tex
\section{Systematic uncertainties and cross-checks}
\label{sec:sys}

The analysis methods are constructed to ensure that systematic biases on the raw
charge asymmetries cancel in the end result. The dominant systematic
uncertainties in both $A_{CP}$ and $A_{CP}|_{S}$ are determined by
considering control decay channels in which no asymmetry is
expected. 

The main systematic uncertainty in $A_{CP}$ results from kinematic
differences between the $\phi\pip$ and $\KS\pip$ final states, which
lead to imperfections in the cancellation of detector asymmetries
between them. Some detector asymmetries arise from small differences in the tracking efficiency or
acceptance across the bending plane of the magnet, i.e. between the
left and right halves of the detector. The response of the hardware trigger is also
known to be asymmetric, because it does not take into account
which way particles bend in the magnetic field when it measures their
transverse energy $E_{\mathrm{T}}$ or \pt. In data taken with
one magnet polarity, a pion from a \Dp decay will bend in the opposite
direction to a pion from a \Dm decay, and if one of these pions is bent
into an inefficient part of the tracking system and is lost while the other
is detected, a charge
asymmetry will result. The same situation could occur if one pion is
bent inwards and so does not meet the hadron trigger $E_{\mathrm{T}}$
threshold while a pion of the opposite charge is bent outwards and therefore
has enough measured $E_{\mathrm{T}}$ to activate the trigger. This cancels to a
good approximation between the left and right halves of the detector,
but any left/right asymmetry in the calorimeters or
muon stations could result in imperfect cancellation, biasing the charge asymmetry. The effect of these asymmetries on this
analysis is not eliminated by the subtraction of the asymmetries in
the two final states in Eq.~\ref{eq:acp} as the two decays do not have
identical kinematic properties. Thus, in the data taken with one magnet
polarity, the charge asymmetry can be affected. However, when the magnet polarity is
reversed, the bias on the asymmetry
changes sign because the particles are deflected in the opposite
directions. The values of $A_{CP}$ in $\Dp \to \phi\pip$ decays are found to differ by
$(0.81\pm0.28)$\% between the data taken with magnet polarity up and data
taken with polarity down. The effect is removed, to a very good approximation, by combining results
obtained with opposite magnet polarities, $A_{CP}^{\uparrow}$ and $A_{CP}^{\downarrow}$, in an average with
equal weights,
\begin{equation}
A_{CP} = \frac{A_{CP}^{\uparrow} + A_{CP}^{\downarrow}}{2}.
\end{equation}

However, non-cancelling effects can bias the measurement and are
considered as sources of systematic uncertainty. The data
triggered by the $\KS$ or $\phi$ meson at the hardware level are
charge-symmetric to a good approximation, and are assumed to be
unbiased. However, in data triggered by another particle in the event,
the particle that activates the trigger may be correlated to
the signal decay. For example, a signal decay is often
accompanied by a $\Dpm$ meson of the opposite charge. If this meson
decays to a charged hadron, electron or muon, the daughter particle,
which is more likely to have the opposite charge to the signal $\Dpm$, could fire the
trigger. The different kinematics and
acceptance of the signal and control channel studied in this analysis mean that
the cancellation of charge-asymmetric trigger efficiencies between them may not be
complete. To study the size of this effect, a sample of approximately
57 million $\Dp \to
\Km\pip\pip$ decays is selected using the same criteria as those for
the signal. The charge asymmetries in the differently triggered datasets
are given in Table~\ref{tab:tiskpp}. Small but significant
discrepancies between data from different triggers are observed,
indicating that the hardware triggers may introduce small biases into
the dataset. The large difference between magnet up and magnet down
data in the sample that is triggered by the muon detectors is due to
a charge-asymmetric \pt threshold in the detector, but this cancels when
the magnet polarities are averaged. A systematic uncertainty equal to the maximum deviation
from the average charge asymmetry of $(-2.034\pm0.014)\%$ in any of the triggers is
assigned. This occurs in the electron trigger and the difference is 0.114\%. The
precision with which effects cancel between $\phi\pip$ and $\KS\pip$ final
states in the analysis cannot be quantified accurately. Therefore the most conservative approach is adopted and no cancellation is assumed.

\begin{table}
\caption{Raw charge asymmetries, in \%, in samples of the $\Dp \to
  \Km\pip\pip$ control decay in which a particle not
  from the signal decay activated various hardware
  triggers. \label{tab:tiskpp}}
\begin{center}
\begin{tabular}{lcccc}
Trigger type & Magnet up & Magnet down & Average & Difference \\
\hline
Hadron & $-2.037\pm0.032$ & $-1.970\pm0.027$ & $-2.003\pm0.021$ & $-0.068\pm0.042$ \\
Muon  & $-2.361\pm0.041$ & $-1.607\pm0.035$ & $-1.984\pm0.027$ & $-0.754\pm0.053$ \\
Electron & $-2.094\pm0.048$ & $-2.201\pm0.041$ & $-2.148\pm0.031$ & $\phantom{-}0.106\pm0.063$ \\
Photon  & $-1.937\pm0.070$ & $-2.230\pm0.060$ & $-2.083\pm0.046$
& $\phantom{-}0.293\pm0.092$ \\
Overall average & $-2.128\pm0.021$ & $-1.940\pm0.018$
&$-2.034\pm0.014$ & $-0.188\pm0.028$
\end{tabular}
\end{center}
\end{table}

Residual detector asymmetry differences between the $\Dp \to \phi\pip$ and $\Dp \to
\KS\pip$ decays due to their different kinematics are studied by
applying several different kinematic binning schemes to the data. The
measured asymmetry is found to be stable with variations in the
binning, suggesting that the detector asymmetries are small. The results are summarised in
Table~\ref{tab:kinbin}. The largest discrepancy in raw asymmetry, as expected,
results from using no kinematic binning, as this does not account
for any variation of the \Dpm production asymmetry across the kinematic
region. The next largest difference with respect to the baseline
binning scheme, of 0.035\%, is assigned as a systematic uncertainty on
the asymmetry due to residual kinematic differences between decay modes.

\begin{table}
\caption{Changes to the final result observed with various alternative
  kinematic binning schemes. The default scheme uses four bins of \DsDp \pt
  and three bins of \DsDp $\eta$. The variable $\phi$ is the azimuthal
  angle around the proton beams. The statistical uncertainties are
  determined by subtracting the uncertainties on the alternative result and the default
  result in quadrature.\label{tab:kinbin}}
\begin{center}
\begin{tabular}{lc}
Binning & Change in $A_{CP}$ ($\times10^{-4}$) \\
\hline
No binning & $\phantom{-}8.3\pm3.7$\\ 
12 bins ($3\times\DsDp$ \pt, $4\times\DsDp$ $\eta$) & $\phantom{-}0.6\pm1.7$\\
48 bins ($8\times \DsDp$ \pt, $6\times \DsDp$ $\eta$) & $-2.9\pm1.1$ \\
192 bins ($2\times \pip$ $p$, $8\times \pip$ $\phi$, $4\times \DsDp$ \pt, $3\times \DsDp$ $\eta$) & $-2.4\pm1.1$\\
180 bins ($3\times \pip$ \pt, $5\times \pip$ $\eta$, $4\times \DsDp$ \pt, $3\times \DsDp$ $\eta$) & $\phantom{-}3.5\pm2.6$ \\
1440 bins ($3\times\pip$ \pt, $5\times\pip$ $\eta$, $8\times\pip$ $\phi$, $4\times\DsDp$ \pt, $3\times\DsDp$ $\eta$) & $\phantom{-}2.5\pm1.6$\\
\end{tabular}
\end{center}

\end{table}
 
The $A_{CP}|_{S}$ observable is highly robust against
systematic uncertainties. Any effect that does not vary
across the Dalitz plot will cancel in the subtraction in Eq.~\ref{eq:acps}, and effects
that do vary with $\Km\pip$ or $\Km\Kp$ invariant mass across the $\phi$
region will also cancel when the regions are combined in the diagonal difference. For
example, the asymmetry  in the interaction of the charged kaons with the
detector material would affect the asymmetry difference between decays
with high and low values of $\Km\pip$ invariant mass, which is correlated with
the momenta of the kaons. However such effects cancel to a
good approximation in both observables, as shown below. Only
quantities that vary between the diagonals of the Dalitz plot region would lead to significant systematic
biases on $A_{CP}|_{S}$. To test for the presence of such effects, $A_{CP}|_{S}$ is calculated in the $\Dsp \to \phi\pip$ control
decay, which has similar kinematics to the signal despite the
different Dalitz plot distributions of the events. The result is
$(-0.120\pm0.119)\%$, which is compatible with zero as expected. The statistical uncertainty on this
result, added in quadrature to the central value, gives a measure of the precision with
which detector effects are known to cancel. Thus a value of 0.169\% is assigned as the main systematic
uncertainty in $A_{CP}|_{S}$. 

The systematic uncertainty due to charged kaon interaction asymmetries is studied by determining the effect
on the result of enlarging the size of the $\Km\Kp$ mass window under
study. This increases the differences between the momentum spectra of
the kaons, which increases the effect of the interaction asymmetry because it depends strongly on
momentum. The consistency of this procedure is checked with simulation studies. The systematic
uncertainty is found to be 0.031\% in $A_{CP}$ for the \Dp decay,
0.002\% for $A_{CP}$ in the \Dsp decay and 0.009\% in $A_{CP}|_{S}$. 

The asymmetric interaction of the neutral kaons with detector
material is studied using the method outlined in
Ref.~\cite{Ko:2010mk} to account for coherent regeneration. The amount of material each kaon passes through before it
decays and the predicted differences between the \Kz and \Kzb material interaction cross
sections~\cite{PhysRevLett.42.13} are used to determine an expected asymmetry. The size of the
effect is found to be $(0.039\pm0.004)\%$, where the uncertainty is due to imperfect
knowledge of the amount of material in the detector. This is consistent with the dependence of
the asymmetry on the depth of material passed through by the kaons seen in
data. The asymmetry is assigned as a systematic uncertainty on the $A_{CP}$
measurements.

A systematic uncertainty of 0.056\% is associated with the resolution in the
Dalitz plot variables for $A_{CP}|_{S}$, due to candidates
migrating across the boundaries of the regions $A-D$. This is determined by taking
the difference between results before and after the $\Dp$ mass is
constrained to the world average value. This procedure is repeated for
$A_{CP}$, but as expected the systematic uncertainty is much smaller.

Further small systematic uncertainties arise from the mass fitting, from the
calculation of the effect of the CPV in the neutral kaon system~\cite{LHCb:2012fb}, from the choice of fiducial cuts, from modelling of
the cross-feed in the $\Dp \to \KS\pip$ decay, and from neglecting the
background in the calculation of $A_{CP}|_{S}$. In the simulation, the contribution of $D$
from $B$ decays is found to differ between the
final states by around 1\%, and this leads to another small uncertainty since the
production asymmetries for $B$ and $D$ decays may differ. 

Other potential sources of systematic uncertainty, such as the difference in selection criteria between the two
final states, are negligible. The kinematic distributions of daughter
particles are checked for biases. The variation of the
asymmetries with time during the data taking period is also considered. The systematic
uncertainties are summarised in Table~\ref{tab:sys}.

\begin{table}
\caption{Systematic uncertainties on the three
  measurements. The
  abbreviation n/a is used where the systematic
  effect does not apply. The row labelled `Backgrounds' represents the
  uncertainty in modelling the cross-feed in $A_{CP}$ and the uncertainty
  from ignoring the background in $A_{CP}|_{S}$.
\label{tab:sys}}
\begin{center}
\begin{tabular}{lccc}

Source & $A_{CP}$ (\Dp) [\%]& $A_{CP}$ (\Dsp) [\%]&
$A_{CP}|_{S}$ [\%]\\
\hline
Triggers & 0.114 & 0.114 & n/a\\
\Dsp control sample size & n/a & n/a & 0.169 \\
Kaon asymmetry & 0.031 &0.002 & 0.009\\
Binning & 0.035 & 0.035  & n/a\\
Resolution  & 0.007 & 0.006 & 0.056 \\
Regeneration & 0.039 & 0.039 & n/a \\
Fitting &0.033 & 0.033 & n/a\\
Kaon \CP violation & 0.028 &0.028 & n/a \\
Fiducial effects & 0.022 & 0.022 & n/a \\
Backgrounds &0.008 &n/a & 0.007\\
$D$ from $B$ & 0.003 & 0.015 & 0.003\\
\hline
Total & 0.138 & 0.136 & 0.178 \\
\end{tabular}
\end{center}

\end{table}

\begin{figure}
\begin{center}
\includegraphics*[width=0.6\columnwidth]{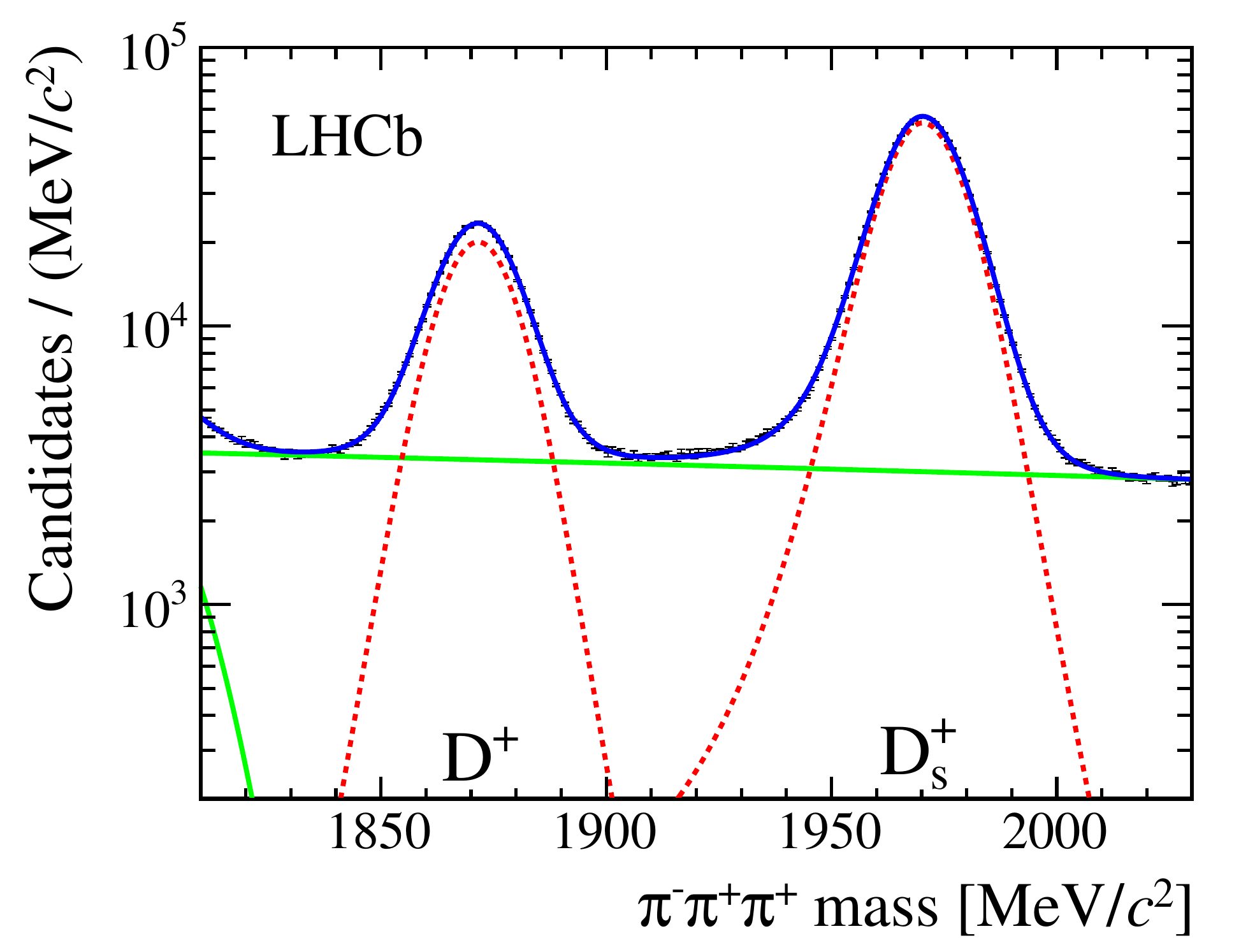}
\end{center}
\caption{Invariant mass distribution of selected
  $\DsDpm \to \pimp\pipm\pipm$ decays. The data are
  represented by symbols with error bars. The red dashed peaks indicate the signal decays,
  the green solid lines represent the combinatorial background shape, and
  the green dotted lines represent backgrounds from mis-reconstructed
 $\Dsp \to \pim\pip\pip\piz$ decays. The blue solid line shows the sum of
 all fit components.\label{fig:3pi}}
\end{figure}

As a further cross check, the difference in raw asymmetry between the $\Dsp \to \phi\pip$ and $\Dsp
\to \pim\pip\pip$ decays is calculated. Since these are both Cabibbo-favoured
tree-level decays, this
quantity is expected to be zero. The $\Dsp \to \pim\pip\pip$ decay has
reasonably similar kinematic properties and a similar yield in our dataset to the $\Dp \to \KS\pip$
decay, and the $\Dsp \to \phi\pip$ is very similar to the
corresponding $\Dp$ decay. Thus the kinematic differences between
the final states in the $\Dsp$ control decays are similar to those in
the $\Dp$ signal channels.

The $\Dsp \to \pim\pip\pip$ decay is reconstructed using
the same selection as for the signal decays. The hardware trigger must
be activated by a particle that does
not form part of the signal decay, or by the \pim meson, or by a random \pip meson. The resulting sample has a large
background due to random pions from the primary vertex. To remove
this, the regions of the $\Dsp \to \pim\pip\pip$ Dalitz plot in which
one of the pions has a low momentum in the \Dsp rest frame are
excluded from the sample by removing the areas of the Dalitz plot
below the $f_{0}(980)$ resonance. The requirement on $\pim\pip$ invariant
mass $m_{\pim\pip}^{2} > 0.75$\gevgevcccc is applied to both
$\pim\pip$ meson pairs. The mass
distribution of the candidates that remain is fitted with a Cruijff
function in the 12 kinematic bins described in Sect.~\ref{sec:yields}
and the raw charge asymmetries in the $\Dsp$ decay are calculated. 

The weighted average of the raw asymmetry differences in the 12
kinematic bins is $(0.22\pm0.12)\%$. The systematic uncertainty on this is
similar to that on the main analysis, or 0.13\%, so the
result differs from zero by 1.3 standard
deviations. This discrepancy is
assumed to be a statistical fluctuation and no additional uncertainty
is assigned.

Many additional cross-checks and comparisons of the data samples are
performed. The raw asymmetries are consistent with
those observed in the measurements of the \Dp and \Dsp production
asymmetries~\cite{LHCb:2012fb, Aaij:2012cy}. The different triggers used in the analysis
give statistically compatible results. A study of
the values of $A_{CP}$ in individual bins gives no
indication of any dependence on \pt and $\eta$. The regions $A-D$ used
in the calculation of $A_{CP}|_{S}$ have fully
compatible asymmetries. 

%% file: results.tex
\section{Results and conclusion}
\label{sec:res}

Searches for \CP violation in the $\phi$ region of the $\Dp \to \Km\Kp\pip$
Dalitz plot and in the $\Dsp \to
\KS\pip$ decay mode are
performed. The results are
\begin{align*}
A_{CP}(\Dp \to \phi\pip) &= (-0.04\pm0.14\pm0.14)\%, \\
A_{CP}|_{S}(\Dp \to \phi\pip)  &= (-0.18\pm0.17\pm0.18)\%, \\ 
A_{CP}(\Dsp \to \KS\pip) &= (+0.61\pm0.83\pm0.14)\%,
\end{align*}
consistent with existing measurements. The first and third
measurements assume negligible \CP violation effects in the $\Dp \to
\KS\pip$ and $\Dsp \to \phi\pip$ control channels, respectively. The $A_{CP}|_{S}$
observable is shown to increase the sensitivity of the analysis to
certain types of \CP violation significantly, but there is no evidence
for \CP violation in either decay. This is the most precise analysis
of \CP violation in the $\phi$ region of the $\Dp \to \Km\Kp\pip$ Dalitz plot
to date. The results suggest that any CP
asymmetries in decays within this region are unlikely to exceed the approximate level
of effects currently believed to be possible within the Standard Model.

%% file: acknowledgements.tex
\section*{Acknowledgements}

\noindent We express our gratitude to our colleagues in the CERN
accelerator departments for the excellent performance of the LHC. We
thank the technical and administrative staff at the LHCb
institutes. We acknowledge support from CERN and from the national
agencies: CAPES, CNPq, FAPERJ and FINEP (Brazil); NSFC (China);
CNRS/IN2P3 and Region Auvergne (France); BMBF, DFG, HGF and MPG
(Germany); SFI (Ireland); INFN (Italy); FOM and NWO (The Netherlands);
SCSR (Poland); ANCS/IFA (Romania); MinES, Rosatom, RFBR and NRC
``Kurchatov Institute'' (Russia); MinECo, XuntaGal and GENCAT (Spain);
SNSF and SER (Switzerland); NAS Ukraine (Ukraine); STFC (United
Kingdom); NSF (USA). We also acknowledge the support received from the
ERC under FP7. The Tier1 computing centres are supported by IN2P3
(France), KIT and BMBF (Germany), INFN (Italy), NWO and SURF (The
Netherlands), PIC (Spain), GridPP (United Kingdom). We are thankful
for the computing resources put at our disposal by Yandex LLC
(Russia), as well as to the communities behind the multiple open
source software packages that we depend on.